\documentclass[epj, twocolumn,nopacs]{svjour}
\usepackage{amssymb,amsmath}
\usepackage{hyperref,graphicx}
\usepackage{dcolumn}
\usepackage{color}
\usepackage{ulem}
\usepackage{bbold}
\begin{document}

\newcommand{\bs}[1]{{\boldsymbol{#1}}}
\newcommand{\bk}{\bs{k}}
\newcommand{\bq}{\bs{q}}
\newcommand{\br}{\bs{r}}
\newcommand{\be}{\bs{\epsilon}}
\newcommand{\bg}{\overline{\bs{G}}_{\omega_1}}
\newcommand{\bgs}{\overline{\bs{G}}_{\omega_2}^*}
\newcommand{\av}[1]{\overline{#1}}
\newcommand{\rmd}{\mathrm{d}}
\newcommand{\Tr}{\mathrm{Tr}}
\newcommand{\intdtwo}[1]{\int \frac{\rmd^2 #1}{(2\pi)^2}}
\newcommand{\intdones}[1]{\int \rmd^2 #1}
\newcommand{\intdthree}[1]{\int \rmd^3 #1}
\newcommand{\intdone}[1]{\int_0^\infty \frac{\rmd #1}{2\pi}} 
\newcommand{\new}[1]{\textcolor{red}{#1}}
\newcommand{\old}[1]{\textcolor{blue}{\sout{#1}}}
\newcommand{\comm}[1]{\textcolor{magenta}{#1}}
\renewcommand\arraystretch{2.2}

\title{Fluctuations of the Casimir potential above a disordered medium}

\author{Nicolas Cherroret \and 
Romain Gu\'erout \and
Astrid Lambrecht \and
Serge Reynaud 
\thanks{\emph{Corresponding author:} cherroret@lkb.upmc.fr}
}  
\institute{
\mbox{Laboratoire Kastler Brossel, UPMC-Sorbonne Universit\'es, CNRS, ENS-PSL Research University, Coll\`{e}ge de France,}\\
\mbox{4 Place Jussieu, 75005 Paris, France}
}

\date{Received: date / Revised version: date}

\abstract{
We develop a general approach to study the statistical fluctuations of the Casimir potential felt by an atom approaching a dielectric disordered medium. Starting from a microscopic model for the disorder, we calculate the variance of potential fluctuations in the limit of a weak density of heterogeneities. We show that fluctuations are essentially governed by scattering of the radiation on a single heterogeneity, and that they become larger than the average value predicted by effective medium theory at short distances. Finally, for denser disorder we show that multiple scattering processes become relevant. 
}

\maketitle

\section{Introduction}
When approached close to each other, two materials experience an attractive Casimir force due to quantum vacuum fluctuations \cite{Casimir48}. In the context of atom-surface interaction, a careful description of the Casimir-Polder effect \cite{CasimirPolder48} is of paramount importance for quantum reflection of cold atoms from surfaces \cite{Pasquini04}, single-atom manipulation on microchips \cite{McGuirk04,Lin04} or trapping of antimatter \cite{Voronin11,Voronin12} to cite a few examples. In all these cases, the Casimir force is usually the dominant one in a short-distance domain typically ranging from hundreds of nanometers to a few micrometers, where possible electrostatic forces are negligible \cite{Naji10}.
In general, the essential features of the Casimir interaction between an atom and a surface are correctly captured by an effective medium description where all the material heterogeneities are averaged out, so that radiation is reflected specularly \cite{Dufour13a}. In real systems however, specular reflection is always an idealization. Some part of electromagnetic radiation is scattered in a more or less complicated way by the material and is eventually reflected in any direction, giving rise to a non-specular contribution to the Casimir interaction potential \cite{Lambrecht06}. For very efficient specular reflectors such as mirrors, the non-specular part of radiation is of course very small. But for strongly heterogeneous systems such as nanoporous materials, powders, or more generically \emph{disordered media}, the contribution of non-specular reflection may be non-negligible and lead to significant fluctuations of the potential around the prediction of effective medium theory. This statement is especially true for dilute disordered media that contain a large fraction of vacuum, such that the effective dielectric constant is close to one and the Casimir potential becomes small. The crucial question that we address in the present paper is then to know whether the Casimir potential may become even smaller than its non-specular fluctuations. In the context of quantum reflection of cold atoms on Casimir potentials \cite{Dufour13b}, a positive answer could explain the low values of reflection coefficients observed in recent experiments using heterogeneous materials \cite{Pasquini06}, as stemming from atoms reflected in non-specular directions.

In order to achieve this goal, we develop in this paper a general description of the Casimir potential between an atom and a heterogeneous material, combining techniques from both the theory of disordered systems \cite{Lagendijk96,Rossum99} and the scattering approach to Casimir forces \cite{Lambrecht06,Emig07}. We consider a generic microscopic model where an atom interacts with a disordered dielectric material consisting of a large collection of heterogeneities (``scatterers'') embedded in a homogeneous background. We describe this system by means of a statistical approach, assuming that the positions of the scatterers are randomly distributed in the material (sec. \ref{framework}). From this model, we first evaluate the ensemble average Casimir potential, and recover the prediction of effective medium theory, which describes a disordered material by a homogeneous dielectric constant (sec. \ref{effective_medium_section}). Then, in sec. \ref{fluctuations_section} we calculate the statistical fluctuations of the potential due to non-specular reflection on the heterogeneities of the material. The results obtained in that section constitute the core of our work, and allow us to provide a rigorous quantification of the role of heterogeneities on the Casimir potential between an atom and a disordered medium. Finally, in sec. \ref{double_scattering} we demonstrate that for a dilute disordered medium, non-specular fluctuations of the Casimir potential are essentially governed by scattering of radiation on a single heterogeneity, whereas for denser disorder multiple scattering processes become significant.

\section{Framework and hypotheses}
\label{framework}

\begin{figure}
\centering
\includegraphics[width=0.7\linewidth]{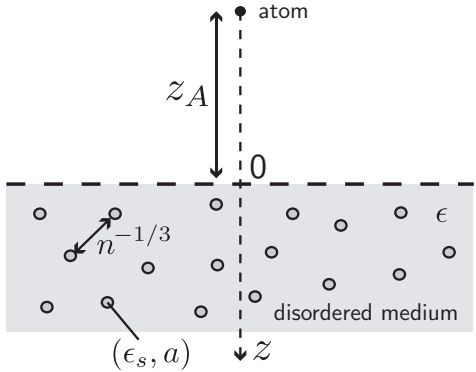}
\caption{
\label{scheme}
We study the Casimir interaction potential between a ground-state atom (placed in vacuum) and a semi-infinite disordered medium. The disordered medium consists of a collection of scatterers (size $a$, relative dielectric constant $\epsilon_s$, density $n$) whose positions are uniformly distributed in a homogeneous background of relative dielectric constant $\epsilon$. We assume a flat interface between the medium and the vacuum.
}
\end{figure}

We consider a ground-state, two-level atom in vacuum, located at distance $z_A>0$ from a semi-infinite disordered medium, as shown in fig. \ref{scheme}. The response of the atom to an electric field of frequency $\omega$ is characterized by a simple model for the dynamic polarizability $\alpha(\omega)=\alpha(0)\omega_A^2/(\omega_A^2-\omega^2)$, where $\omega_A$ is the atomic resonance frequency (here and in the rest of the paper, polarizabilities are expressed in SI units divided by $\epsilon_0$). While the two-level atom approach may fail in general to describe accurately the dispersion interaction of real atoms with a surface at short distances \cite{Barton74}, it has been shown to be a good description for the Casimir-Polder interaction of a nanosphere and a surface \cite{Canaguier-Durand11}.
The disordered medium is assumed to be a heterogeneous dielectric material, consisting of a collection of scatterers (size $a$, relative dielectric constant $\epsilon_s$, density $n$) embedded in a homogeneous background of relative dielectric constant $\epsilon>1$, see fig. \ref{scheme}. 

In order to evaluate the Casimir interaction potential $U(z_A)$ between the atom and the disordered medium, a convenient approach is the scattering formalism \cite{Lambrecht06}, here written at zero temperature and in the dipolar approximation for the atom \cite{Emig07,Messina09}:
\begin{eqnarray}
\label{Udef}
U(z_A)&=&-\dfrac{\hbar}{c^2}\text{Im}\bigg[\intdone{\omega}
\int\dfrac{\rmd^2\bq_a}{(2\pi)^2}
\dfrac{\rmd^2\bq_b}{(2\pi)^2}
\sum_{p_a, p_b} i\omega^2 \alpha(\omega)\nonumber\\
&&\times
r_{ab}(\omega)
\dfrac{e^{i(k_a^z+k_b^z)z_A}}{2k_{a}^z} 
\be_{p_a}(\bq_a)\cdot\be_{p_b}(\bq_b)\bigg].
\end{eqnarray}
In eq.~(\ref{Udef}), $r_{ab}(\omega)\equiv\left<\bq_b, p_a|\br(\omega)|\bq_a,p_a\right>$ is the reflection coefficient describing the scattering of an incoming mode with transverse wave vector $\bq_a$ and polarization vector $\be_{p_a}(\bq_a)$ ($p_a=$transverse electric TE, transverse magnetic TM) into an outgoing mode with transverse wave vector $\bq_b$ and polarization $p_b$, at frequency $\omega$ (the frequency dependence of polarization vectors will be generally omitted). Its computation requires the knowledge of the reflection tensor $\br(\omega)$ of the disordered medium. The two exponential factors $e^{ik_a^z z_A}$ and $e^{ik_b^z z_A}$ respectively account for the propagation of these modes from the atom to the disordered medium, and from the disordered medium to the atom, with longitudinal wave numbers $k_a^z=\sqrt{\omega^2/c^2-\bq_a^2}$ and $k_b^z=\sqrt{\omega^2/c^2-\bq_b^2}$. The Casimir potential is eventually obtained by summing over all incoming and outgoing modes and over all frequencies. 

In this paper, we make use of a statistical description of the disordered medium. This means that the reflection coefficient is considered as a \emph{random} quantity $\br=\av{\br}+\delta\br$, characterized by an average value, $\av{\br}$, and by fluctuations, $\delta\br$, giving rise to an average potential $\av{U}(z_A)$ and to potential fluctuations $\delta U(z_A)$, respectively. In general, the ensemble average $\av{(\dots)}$ over the statistics of the disorder can be very difficult to perform. Indeed, the scatterers can be spatially organized according to a more or less complex pattern. They can also have a complicated internal structure with many resonances, and possibly a distribution of sizes (polydispersity). In order to present a scenario as simple as possible, in this paper we choose a statistical model where all heterogeneities are identical, Rayleigh scatterers (i.e. of size $a\ll\lambda$), and where the position $\br_i$ of each scatterer follows a uniform distribution (the so-called ``Edwards model'' \cite{Edwards58}). With these assumptions, the ensemble average simply amounts to summing over the positions of scatterers: $\overline{(\ldots)}\equiv\prod_i\int(\rmd\br_i/\Omega)(\ldots)$, where $\Omega$ is the volume of the system and the sum is over the total number $N$ of scatterers. We consider here the thermodynamic limit $\Omega\rightarrow\infty$, $N\rightarrow\infty$, with a constant density of scatterers, $n=N/\Omega$. 
Finally, we restrict ourselves to a dilute disordered medium, for which the distance $n^{-1/3}$ between the scatterers is large compared to their typical size $a$:
\begin{equation}
\label{diluteness}
n a^3\ll 1.
\end{equation}
Such a concentration is typically encountered in porous materials, where $n a^3$ can be down to a few percents \cite{Sinko10,Granitzer10}. The physical consequences of diluteness on the fluctuations of the Casimir potential will be discussed in sec. \ref{double_scattering}.

\section{Average Casimir potential: effective medium theory}
\label{effective_medium_section}

In this section, we evaluate the average Casimir potential $\av{U}(z_A)$, starting from the microscopic model of disorder introduced in sec. \ref{framework}. This calculation will allow us to recover known results from Casimir physics,
as well as to introduce the necessary theoretical tools for the description of fluctuations presented in sec. \ref{fluctuations_section}.

\subsection{Preliminary: average Green tensor}

Before calculating $\av{U}(z_A)$, let us introduce a convenient tool to describe a disordered system, the Green tensor $\bs{G}_\omega(\br,\br')$, which is solution of the Helmholtz equation
\begin{equation}
\label{Helmholtz}
\left[k^2\epsilon(\br)-\bs{\nabla}\times\bs{\nabla}\times\right]\bs{G}_\omega(\br,\br')=\delta{(\br-\br')}\bs{I},
\end{equation}
where $k=\omega/c$ and $\bs{I}$ denotes the unit tensor of rank 2. Let us first leave aside the geometry of fig. \ref{scheme} for a while, and consider the case of an infinite disordered medium described by the Edwards model, i.e. with
\begin{equation}
\label{epsilonr}
\epsilon(\br)=\epsilon\left[1+\sum_i v(\br-\br_i)\right],
\end{equation}
where $v(\br-\br_i)$ represents the (central) potential of an individual scatterer, located at point $\br_i$ \cite{Edwards58}. With the assumptions discussed in sec. \ref{framework}, the ensemble average Green tensor can be calculated from scattering theory \cite{Lagendijk96,Rossum99}. We will not reproduce this calculation here but simply give the final result which, for Rayleigh scatterers ($a\ll\lambda=2\pi/\omega$), turns out to be independent of the particular shape of the function $v(\br-\br_i)$:
\begin{equation}
\label{Gbar_inf}
\av{\bs{G}}_\omega(\br,\br')=
-\left(\bs{I}+\dfrac{1}{k^2}\bs{\nabla}\otimes\bs{\nabla}\right)\dfrac{e^{i \tilde{k} |\br-\br'|}}{4\pi |\br-\br'|},
\end{equation}
where $\otimes$ denotes the outer product. In eq. (\ref{Gbar_inf}), information on the disordered nature of the material is contained in the effective wave number $\tilde{k}=k\sqrt{\tilde\epsilon}=k\sqrt{\epsilon(1+n\alpha_s)}$, where $n$ the density of scatterers and $\alpha_s=3u(\epsilon_s-\epsilon)/(\epsilon_s+2\epsilon)$ the static polarizability of a scatterer of volume $u\propto a^3$.
The physical content of eq. (\ref{Gbar_inf}) is that \emph{on average}, the disordered medium can be described as homogeneous, with a relative dielectric constant $\tilde{\epsilon}=\epsilon(1+n\alpha_s)$. This is the so-called \emph{effective medium theory}. Note that in practice, this description amounts to replacing the dielectric constant $\epsilon(\br)$ in the Helmholtz equation (\ref{Helmholtz}) by its average value $\overline{\epsilon(\br)}=\epsilon[1+\overline{\sum_i v(\br-\br_i)}]$. This is easily seen for Rayleigh scatterers, which can be considered point-like ($a\ll\lambda$): $v(\br-\br_i)\simeq\alpha_s\delta(\br-\br_i)$. Then, using the definition of the disorder average given in Sec. \ref{framework} we have\begin{equation}
\av{\epsilon(\br)}=\epsilon\left[1+\int \prod_j\dfrac{\rmd \br_j}{\Omega}\sum_{i=1}^N \alpha_s \delta(\br-\br_i)\right]=\tilde{\epsilon}.
\end{equation}
It should be noted that the effective dielectric constant $\tilde{\epsilon}$ discussed here is frequency independent, which is a direct consequence of our model of point scatterers. In general, the polarizability of the scatterers can have a more complicated frequency dependence with real and imaginary parts, for instance of the type $\alpha_s/(1-\omega^2/\omega_s^2+i\omega^3/\gamma_s)$ for a single resonance \cite{Lagendijk96}. As discussed below though, accounting for such a general dispersion relation would only affect the prefactor of the average Casimir potential at short distances while this would have no effect on its relative fluctuations. For this reason and for the sake of simplicity, we restrict our discussion to the case $\tilde{\epsilon}=\text{const}$, keeping in mind that the quantitative description of a specific material would require a proper modification of $\alpha_s$.

\subsection{Average Casimir potential}
\label{avCP}

Let us now come back to the geometry of fig. \ref{scheme}, where $\epsilon(\br)=1$ for $z<0$ and $\epsilon(\br)$ is given by eq. (\ref{epsilonr}) for $z>0$. Assuming a source point $\br'$ inside the semi-infinite space $z'<0$, we can express the ensemble average Green tensor at a point $\br$ in the disordered medium as \cite{Rossum99}
\begin{equation}
\label{G_semi}
\av{\bs{G}}_\omega(\br,\br')=
     \begin{cases}
        \bs{G}_\omega^{(0)}(\br,\br')+\av{\bs{G}}_\omega^{(r)}(\br,\br') & z<0 \\
        \av{\bs{G}}_\omega^{(t)}(\br,\br') & z>0,
     \end{cases}
\end{equation}
where $\bs{G}_\omega^{(0)}$ is the free-space Green tensor, and $\av{\bs{G}}_\omega^{(r)}$ and $\av{\bs{G}}_\omega^{(t)}$ are components resulting from the reflection and transmission of the incoming wave at the interface. 

As seen in eq. (\ref{Udef}), the calculation of $\av{U}(z_A)$ requires the knowledge of the average reflection coefficient, $\av{r}_{ab}(\omega)$, which describes the scattering from an incoming mode ($\bq_a, p_a)$ into an outgoing mode $(\bq_b, p_b)$. This quantity is related to the Green tensor through \cite{Feng94}
\begin{equation}
\label{rdef}
\av{r}_{ab}(\omega)=2i k_a^z\langle p_b|\av{\bs{G}}_\omega^{(r)}(\{\bq_a, 0\}, \{\bq_b, 0\})| p_a\rangle,
\end{equation}
where we have introduced the two-dimensional Fourier transform of $\av{\bs{G}}_\omega^{(r)}(\br_a,\br_b)\equiv \av{\bs{G}}_\omega^{(r)}(\{\bs{\rho}_a, z_a\}, \{\bs{\rho}_b, z_b\})$ \cite{Feng94}:
\begin{eqnarray}
\av{\bs{G}}_\omega^{(r)}(\{\bq_a, z_a\}, \{\bq_b, z_b\})&=&
\int \rmd^2\bs{\rho}_a
\rmd^2\bs{\rho}_b e^{i \bq_a\cdot\bs{\rho}_a-i \bq_b\cdot\bs{\rho}_b}\nonumber\\
&& \times
\av{\bs{G}}_\omega^{(r)}(\{\bs{\rho}_a, z_a\}, \{\bs{\rho}_b, z_b\}).
\end{eqnarray}

By requiring that the general form (\ref{G_semi}) should be solution of the Helmholtz equation (\ref{Helmholtz}) in the effective medium and imposing, say, the continuity of the transverse component of the electric and magnetic fields at the interface \cite{Chew95}, we readily obtain
\begin{equation}
\label{rav}
\av{r}_{ab}(\omega)=(2\pi)^2\delta(\bq_a-\bq_b)\delta_{p_ap_b}r_{p_a}(\omega),\newline
\end{equation}
where $r_{p_a}(\omega)$ are the usual Fresnel coefficients
\begin{equation}
\label{Fresnel}
r_\text{TE}(\omega)=\dfrac{k_a^z-\tilde{k}_a^z}{k_a^z+\tilde{k}_a^z},\ \ \ \
r_\text{TM}=\dfrac{\tilde{\epsilon}\, k_a^z-\tilde{k}_a^z}{\tilde{\epsilon}\,k_a^z+\tilde{k}_a^z},
\end{equation}
with $\tilde{k}_a^z=\sqrt{\tilde{\epsilon}\,\omega^2/c^2-\bq_a^2}$. The two-dimensional Dirac del-ta function that appears in eq. (\ref{rav}) signals that translation invariance along the transverse directions $x$ and $y$ is recovered after averaging over the positions of the scatterers. In other words, \emph{reflection is specular on average}. The presence of heterogeneities in the medium only manifests itself as an increase of the macroscopic dielectric constant, which becomes $\tilde{\epsilon}=\epsilon(1+n\alpha_s)$ instead of $\epsilon$ in the absence of disorder.

Inserting Eqs. (\ref{rav}) and (\ref{Fresnel}) into eq. (\ref{Udef}), we obtain the average Casimir potential. Using the fact that $\av{r}_{ab}(\omega)$ has no poles in the upper complex sheet due to causality, we can transform the integral over frequencies in a usual way, by performing the Wick rotation $\omega=i\xi$ \cite{Lambrecht06}:

\begin{eqnarray}
\label{avU_final}
\av{U}(z_A)=\dfrac{\hbar}{c^2}\intdone{\xi}\xi^2\alpha(i\xi)\intdtwo{\bq} \dfrac{e^{-2\kappa z_A}}{2\kappa}\nonumber\\
\times\left[r_\text{TE}(i\xi)-\left(1+\dfrac{2c^2 \bq^2}{\xi^2}\right)r_\text{TM}(i\xi)\right],
\end{eqnarray}
\begin{figure}
\includegraphics[width=1\linewidth]{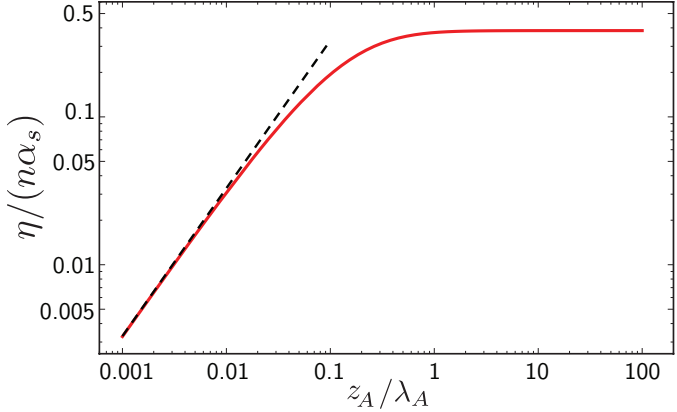}
\caption{
\label{av_potential}
$z_A$-dependence of the reduction factor $\eta=\av{U}(z_A)/U_*$, in units of $n\alpha_s$. The dashed line corresponds to the short-distance asymptotic limit $z_A\ll\lambda_A$, given by eq. (\ref{Ubar}).
}
\end{figure}
where $\kappa=\sqrt{\xi^2/c^2+\bq^2}$. From eq. (\ref{avU_final}), we recover the Casimir potential between an atom and a perfect mirror in the limit $\epsilon\rightarrow\infty$ \cite{Marinescu97,Friedrich02,Voronin05}. 
We recall its behavior at large distances $z_A\gg\lambda_A=2\pi c/\omega_A$, which will be used for comparison in the following:
\begin{equation}
\label{Ustar}
U_*=-\dfrac{3\alpha(0)\hbar c}{32\pi^2 z_A^4}\ \ \ (\epsilon\rightarrow\infty, z_A\gg\lambda_A).
\end{equation}
Of course, when $\epsilon\rightarrow\infty$ radiation is totally reflected from the interface and the disorder underneath plays no role. From here on, we rather focus on the opposite limit $\epsilon\rightarrow 1$ where  reflection purely stems from the effective part $n\alpha_s$ of the dielectric constant. We show in fig. \ref{av_potential} the ratio $\eta\equiv\av{U}(z_A)/U_*$ for this case, in units of $n\alpha_s$ ($\eta$ measures the reduction of the potential with respect to the case of perfect mirrors). For a dilute disordered medium, see eq. (\ref{diluteness}), $n\alpha_s\sim na^3\ll1$, and thus $\av{U}(z_A)\ll U_*$. The $z_A$-dependence of $\av{U}(z_A)$ is, on the other hand, the same as the one obtained for a perfect mirror, i.e. characterized by a qualitatively different asymptotic behavior at short and large distances where retardation effects become significant \cite{Casimir48}:

\begin{equation}
\label{Ubar}
\eta=
     \begin{cases}
        \dfrac{23}{60}n\alpha_s & z_A\gg\lambda_A \\
        \dfrac{\pi^2}{3}\dfrac{z_A}{\lambda_A}n\alpha_s & z_A\ll\lambda_A.
     \end{cases}
\end{equation}
At this stage we recall that Eq. (\ref{Ubar}) has been obtained for scatterers with polarizability $\alpha_s=\text{const}$. For a frequency-dependent polarizability, Eq. (\ref{Ubar}) would be slightly modified at small distances (typically by a constant prefactor of the order of unity) \cite{Casimir48}, but not at large distances where retardation effects select only the static component of $\alpha_s$.

$\av{U}(z_A)=\eta U_*$ is the specular part of the Casimir potential  between the atom and the disordered medium in the limit $\epsilon=1$. Due to the factor $n\alpha_s\ll 1$, this quantity is much smaller than $U_*$ for a dilute distribution of heterogeneities. This is the typical situation where the fluctuations of the potential, originating from non-specular reflection, are likely to play a very important role, as we discuss now.

\section{Fluctuations of the Casimir potential}
\label{fluctuations_section}

Having obtained the average value of the Casimir potential, we now turn to the primary subject of the present work, the study of fluctuations.

\subsection{Diagrammatic approach}

From now on we neglect reflection at the interface, focusing on the limit $\epsilon=1$ where the average Casimir potential is given by eq. (\ref{Ubar}). In order to characterize the fluctuations around $\av{U}(z_A)$, we express $r_{ab}(\omega)=\av{r}_{ab}(\omega)+\delta r_{ab}(\omega)$ in terms of an average value and a fluctuating part. Squaring eq. (\ref{Udef}) and applying the disorder average, we obtain after some algebra 
\begin{equation}
\av{U^2}(z_A)=\av{U}^2(z_A)+\av{\delta U^2}(z_A).
\end{equation}
Here $\av{U}(z_A)$ is the average Casimir potential (\ref{avU_final}), and the variance $\av{\delta U^2}(z_A)$, characterizing fluctuations, is given by \cite{footnote}
\begin{eqnarray}
\label{deltaUdef}
&&\av{\delta U^2}(z_A)=\dfrac{\hbar^2}{c^4}
\text{Re}\bigg[
\int_0^\infty\dfrac{\rmd\omega_1}{2\pi}
\int\dfrac{\rmd^2\bq_a}{(2\pi)^2}
\dfrac{\rmd^2\bq_b}{(2\pi)^2}
\dfrac{\rmd^2\bq_c}{(2\pi)^2}
\dfrac{\rmd^2\bq_d}{(2\pi)^2}
\nonumber\\
&&\sum_{p_a, p_b, p_c, p_d} \omega_1^2\omega_2^2
\alpha(\omega_1)\alpha^*(\omega_2)
\dfrac{e^{i(k_a^z+k_b^z-k_c^{z*}-k_d^{z*})z_A}}{4k_{a}^zk_{c}^{z*}} 
\nonumber\\
&&\times 
\left(\be_a\cdot\be_b\right)
\left(\be_c^*\cdot\be_d^*\right)
\av{\delta r_{ab}(\omega_1)\delta r_{cd}^*(\omega_2)}
\bigg],
\end{eqnarray}
where we have used the short notation $\be_i\equiv\be_{p_i}(\bq_i)$ for $i=a,b,c,d$, and where ``$*$'' denotes complex conjugation.
At this stage, the whole difficulty lies in the evaluation of the correlation function of the fluctuations of the reflection coefficient, $\av{\delta r_{ab}(\omega_1)\delta r_{cd}^*(\omega_2)}$. 
\begin{figure}
\centering
\includegraphics[width=0.8\linewidth]{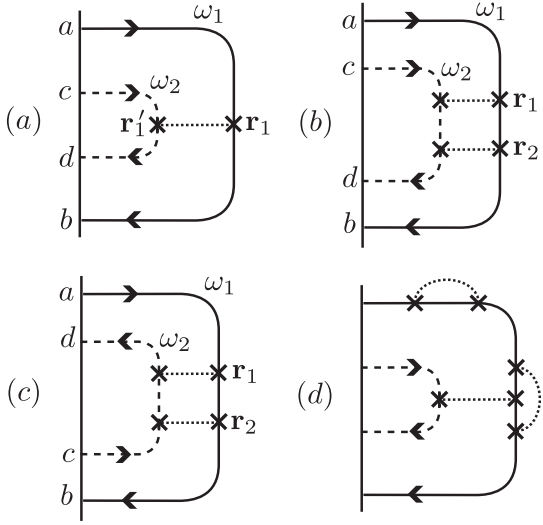}
\caption{Single scattering (a), incoherent double scattering (b) and coherent double scattering (c) contributions to the correlation function of the fluctuations of the reflection coefficient, $\av{\delta r_{ab}(\omega_1)\delta r_{cd}^*(\omega_2)}$. Diagram (d) shows a typical contribution involving recurrent scattering, and is negligible for a dilute disordered medium. Arrows indicate the direction of wave propagation. Solid and dashed lines denote the ensemble average Green tensors $\bg$ and $\bgs$, respectively (scattering processes along these individual paths are not shown explicitly). Vertices consisting of two crosses located at $\bs{r}_1$ and $\bs{r}_1'$ and connected by a dotted line refer to the correlation function $U(\br_1-\br_1')=n(\alpha_sk_1 k_2)^2\delta(\br_1-\br_1')$.}
\label{diagrams}
\end{figure}
According to eq. (\ref{rdef}), $\delta r_{ab}(\omega)=2i k_a^z\langle p_b|\delta\bs{G}_\omega(\{\bq_a, z_a=0\}, \{\bq_b, z_b=0\})| p_a\rangle$, where $\delta \bs{G}_\omega\equiv\bs{G}_\omega-\av{\bs{G}}_\omega$. Therefore, this correlation function is controlled by pairs of wave paths (associated with the Green tensors $\bs{G}_{\omega_1}$ and  $\bs{G}_{\omega_2}^*$) sharing one or several scattering processes. The simplest of these contributions is the one shown in fig. \ref{diagrams}(a): two scattering amplitudes entering the medium in the modes $a=(\bq_a, p_a)$ and $c=(\bq_c, p_c)$ propagate independently in the effective medium at frequency $\omega_1$ and $\omega_2$ respectively (solid and dashed lines), until they encounter a common heterogeneity at points $\br_1$ and $\br_1'$, from which they are scattered. After this process, both amplitudes again propagate independently in the effective medium, and finally leave the material in the modes $b=(\bq_b, p_b)$ and $d=(\bq_d, p_d)$. In what follows, we will refer to the diagram in fig. \ref{diagrams}(a) as the ``single scattering'' contribution to $\av{\delta r_{ab}(\omega_1)\delta r_{cd}^*(\omega_2)}$. It should however be noted that this terminology simply means that the two paths are \emph{correlated} via a single scatterer (before and after this process, individual amplitudes can be scattered an arbitrary number of times). The mathematical formulation of this diagram is
\begin{eqnarray}
\label{rcorrelation}
&&\av{\delta r_{ab}(\omega_1)\delta r_{cd}^*(\omega_2)}=4 k_a^z k_c^{z*}
\int 
\rmd^2\bs{\rho}_a
\rmd^2\bs{\rho}_b
\rmd^2\bs{\rho}_c
\rmd^2\bs{\rho}_d
\nonumber\\
&&\int_{z_1>0}\!\!\!\!\!\!\!\rmd^3 \br_1
\int_{z_1'>0}\!\!\!\!\!\!\!\rmd^3 \br_1'
e^{i (\bq_a\cdot\bs{\rho}_a-\bq_b\cdot\bs{\rho}_b-\bq_c\cdot\bs{\rho}_c+\bq_d\cdot\bs{\rho}_d)} 
 U(\br_1-\br_1')
\nonumber\\
&&
\times\left[\be_a\cdot\av{\bs{G}}_{\omega_1}(\{\bs{\rho}_a,0\},\br_1)\cdot\av{\bs{G}}_{\omega_1}(\br_1, \{\bs{\rho}_b,0\})\cdot \be_b\right]
\nonumber\\
&&
\otimes\left[\be_c^*\cdot\av{\bs{G}}_{\omega_2}^*(\{\bs{\rho}_c,0\},\br_1')\cdot\av{\bs{G}}_{\omega_2}^*(\br_1', \{\bs{\rho}_d,0\})\cdot \be_d^*\right].
\end{eqnarray}
In eq. (\ref{rcorrelation}), quantities referring to the same scattering path ($1$ or $2$) are chained via an inner product $``\cdot"$, while quantities referring to two different scattering paths are chained via an outer product $``\otimes"$. $U(\br_1-\br_1')$ is the correlation function of the fluctuations of the disorder potential $-k^2\sum_i v(\br-\br_i)$ at two different frequencies $\omega_1$ and $\omega_2$. For independent Rayleigh scatterers, $U(\br_1-\br_1')\simeq n(\alpha_s k_1 k_2)^2 \delta(\br_1-\br_1')$, with $k_1=\omega_1/c$ and $k_2=\omega_2/c$ \cite{Lagendijk96,Rossum99}. Furthermore, since we assume no internal reflection of the propagating waves at the interface ($\epsilon=1$), the average Green tensors are simply given by eq. (\ref{Gbar_inf}). Re-expressing them in terms of a Fourier integral, we have for instance
\begin{eqnarray}
\label{rr_intermediate}
&&\av{\bs{G}}_{\omega_1}(\{\bs{\rho}_a,0\}, \br_1)=\nonumber\\
&&\intdtwo{\bq}\dfrac{e^{i\bq\cdot(\bs{\rho}_1-\bs{\rho}_a)+i\tilde{k_a^z} z_1}}{2ik_a^z}
\sum_p\bs{\epsilon}_{p}(\bq)\otimes \bs{\epsilon}_p(\bq),
\end{eqnarray}
where $\tilde{k_a^z}=\sqrt{\tilde{\epsilon}\, \omega_1^2/c^2-\bq_a^2}$. In the limit (\ref{diluteness}), one can safely replace $\tilde{\epsilon}$ by $\epsilon=1$ in all wave numbers, and thus replace $\tilde{k_i^z}$ by $k_i^z$ for all $i=a,b,c,d$. Then, inserting eq. (\ref{rr_intermediate}) into eq. (\ref{rcorrelation}) and performing all spatial integrations, we obtain
\begin{eqnarray}
\label{rr_final}
&&\av{\delta r_{ab}(\omega_1)\delta r_{cd}^*(\omega_2)}=
\dfrac{(2\pi)^2\delta(\Delta\bq_a-\Delta\bq_b)}{4 k_b^z k_d^{z*}(k_c^{z*}-k_a^z-k_b^z+k_d^{z*})}\nonumber\\
&&\times
n k_1^2 k_2^2\alpha_s^2
\left(\be_a\cdot\be_b\right)\left(\be_c^*\cdot\be_d^*\right),
\end{eqnarray}
where $\Delta\bq_a=\bq_a-\bq_c$ and $\Delta\bq_b=\bq_b-\bq_d$. The Dirac delta function is a manifestation of the so-called ``memory effect'': a disordered medium keeps the memory of the direction of an incoming radiation when the latter is changed by a small angle \cite{Freund88,Feng88}.

The last step consists in inserting eq. (\ref{rr_final}) into eq. (\ref{deltaUdef}) and computing the integrals over frequencies and momenta. As for the average potential $\av{U}(z_A)$, this calculation is strongly facilitated by the application of a Wick rotation in the frequency domain. This procedure however deserves a comment, as the Wick rotation now involves two frequencies. The treatment of the frequency $\omega_1$ is based on the same reasoning as that of sec. \ref{avCP}: due to the causality, the function $r_{ab}(\omega_1)$ has no poles in the upper complex sheet $\text{Im}(\omega_1)>0$, which guides us to performing the Wick rotation $\omega_1=i\xi_1$. The argument is slightly different for the frequency $\omega_2$, since it is now the conjugate of $r_{cd}$ that is involved in eq. (\ref{deltaUdef}). We can however still appeal to causality by noticing that $r_{cd}^*(\omega_2)=r_{cd}(-\omega_2)$: this function has no poles in the lower complex sheet $\text{Im}(\omega_2)<0$, which now imposes the Wick rotation $\omega_2=-i\xi_2$. This procedure finally leads to 
\begin{eqnarray}
\label{deltaU1_final}
&&\av{\delta U^2}(z_A)=
\dfrac{\hbar^2}{c^8}
\int_0^\infty
\dfrac{\rmd\xi_1}{2\pi}
\dfrac{\rmd\xi_2}{2\pi}
\int 
\dfrac{\rmd^2\bq_a}{(2\pi)^2}
\dfrac{\rmd^2\bq_b}{(2\pi)^2}
\dfrac{\rmd^2\bq_d}{(2\pi)^2}
\xi_1^4\xi_2^4\nonumber\\
&&\times\alpha(i\xi_1)\alpha(i\xi_2)\dfrac{e^{-(\kappa_a+\kappa_b+\kappa_c+\kappa_d)z_A}}{16 \kappa_a\kappa_b\kappa_c\kappa_d}
\dfrac{n\alpha_s^2}{\kappa_a+\kappa_b+\kappa_c+\kappa_d}\nonumber\\
&&\times\sum_{p_a, p_b, p_c, p_d}
\left(\be_a\cdot\be_b\right)^2
\left(\be_c\cdot\be_d\right)^2,
\end{eqnarray}
where $\kappa_a=\sqrt{\xi_1^2/c^2+\bq_a^2}$, $\kappa_b=\sqrt{\xi_1^2/c^2+\bq_b^2}$, and $\kappa_d=\sqrt{\xi_2^2/c^2+\bq_d^2}$, $\kappa_c=\sqrt{\xi_2^2/c^2+(\bq_a-\bq_b+\bq_d)^2}$. The explicit value of the various scalar products $\be_i\cdot\be_j$ is given in Appendix \ref{scalar}. eq. (\ref{deltaU1_final}) cannot be further simplified and has to be evaluated numerically.

\subsection{Results}
\label{results}

\begin{figure}
\includegraphics[width=1\linewidth]{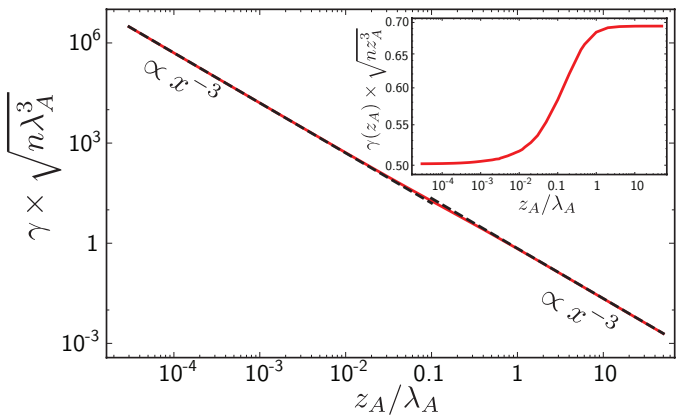}
\caption{
\label{relative}
Main panel: $z_A$-dependence of the relative fluctuations of the Casimir potential, $\gamma$, in units of $1/\sqrt{n\lambda_A^3}$. Dashed lines are the short- and large-distance asymptotic limits given by eq. (\ref{gamma}). Inset: residual $z_A$-dependence of $\gamma\times\sqrt{n z_A^3}$. The curve tends to the constants $b_1\simeq 0.5$ and $a_1\simeq 0.7$ at short and large distances, respectively.}
\end{figure}

Let us introduce the ratio 
\begin{equation}
\label{gamma_def}
\gamma\equiv
\dfrac{\sqrt{\av{\delta U^2}(z_A)}}{|\av{U}(z_A)|}
=\dfrac{\sqrt{\av{\delta U^2}(z_A)}}{\eta |U_*|},
\end{equation}
which measures the single scattering contribution to the relative fluctuations of the Casimir potential. From eq. (\ref{deltaU1_final}), we find the following asymptotic limits:
\begin{equation}
\label{gamma}
\gamma=
     \begin{cases}
      \dfrac{a_1}{\sqrt{n z_A^3}} & z_A\gg\lambda_A \\
      \dfrac{b_1}{\sqrt{n z_A^3}} & z_A\ll\lambda_A,
     \end{cases}
\end{equation}
with $a_1\simeq 0.7$ and $b_1\simeq 0.5$. Several important conclusions can be drawn from this result. First, as for the average potential, $\av{U}(z_A)$, the behavior of fluctuations, $[\av{\delta U^2}(z_A)]^{1/2}$, is qualitatively different at short and at large distances due to retardation effects. Because of the additional factor $(n z_A^3)^{-1/2}$ however,  the $z_A$-dependence of $[\av{\delta U^2}(z_A)]^{1/2}$ is not the same as the one of $\av{U}(z_A)$. Indeed, from Eqs. (\ref{gamma}), (\ref{Ustar}) and (\ref{Ubar}) we find $[\av{\delta U^2}(z_A\gg\lambda_A)]^{1/2}\propto z_A^{-11/2}$ and $[\av{\delta U^2}(z_A\ll\lambda_A)]^{1/2}\propto z_A^{-9/2}$. Second, eq. (\ref{gamma}) suggests that the \emph{relative} fluctuations, $\gamma$, are essentially controlled by the single factor $(n z_A^3)^{-1/2}$, both at short and large distances. This statement is confirmed by fig. \ref{relative}, which displays the relative fluctuations computed from eq. (\ref{deltaU1_final}) for any value of $z_A/\lambda_A$.
The proportionality of $\gamma$ to $(n z_A^3)^{-1/2}$ has a simple physical interpretation: as the atom is approached to the surface, the Casimir potential at distance $z_A$ from the medium is controlled by the interaction of radiation with the matter contained in a volume $\propto z_A^3$. Relative fluctuations are then of the order of $1/\sqrt{N_{z_A}}$, where $N_{z_A}\equiv nz_A^3$ is the number of scatterers in that volume. 
Note that beside the essential dependence in $(n z_A^3)^{-1/2}$, $\gamma$ also varies very slightly with $z_A/\lambda_A$. In eq. (\ref{gamma}), this manifests itself in the two different numerical prefactors $a_1$ and $b_1$. This residual dependence is shown in the inset of fig. \ref{relative}, which displays $\gamma\times\sqrt{n z_A^3}$ as a function of $z_A/\lambda_A$.

Eq. (\ref{gamma}) provides a simple criterion for the relevance of fluctuations in an experiment: fluctuations can only be neglected when $z_A\gg n^{-1/3}$, the typical distance between the scatterers. On the contrary, when $z_A\ll\ n^{-1/3}$, fluctuations become larger than the prediction of effective medium theory and can thus no longer be ignored. Furthermore, although we have considered here a simple model of scatterers for which $\alpha_s=\text{const}$, we have verified that Eq. (\ref{gamma}) remains valid for a more general dispersion relation, even at short distances. Indeed, in that case both $\overline{U}^2$ and $\overline{\delta^2U}$ are modified by the same amount, thus leaving the relative fluctuations unchanged.

\section{Double scattering contribution}
\label{double_scattering}

\begin{figure}
\includegraphics[width=1\linewidth]{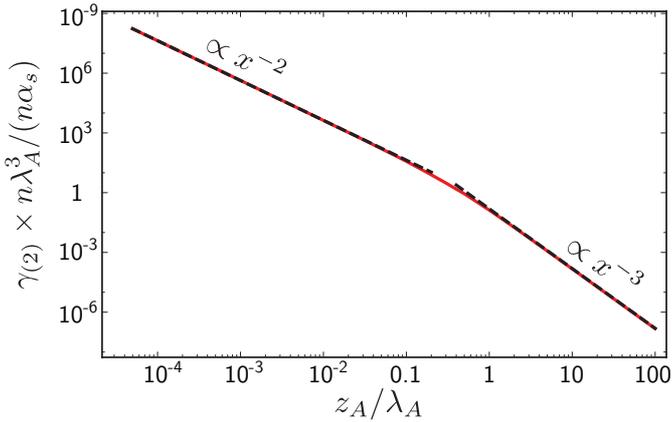}
\caption{
\label{fluc_double}
$z_A$-dependence of the double scattering contribution to the relative fluctuations of the Casimir potential, $\gamma_{(2)}$, in units of $n\alpha_s/(n\lambda_A^3)$. Dashed lines are the short- and large-distance asymptotic limits given by eq. (\ref{gamma2}).}
\end{figure}

In the previous section, we have calculated the single scattering contribution to the fluctuations of the Casimir potential, diagram (a) in fig. \ref{diagrams}. In order to estimate the role played by multiple scattering of light inside the disordered material, we now propose to calculate the contribution due to double scattering. This contribution is characterized by the two processes described by the diagrams (b) and (c) in fig. \ref{diagrams}. In the first one, the two scattering amplitudes share two common heterogeneities, from which they are scattered in the same order (``incoherent contribution''). In the second diagram on the other hand, scattering amplitudes propagate in opposite directions (``coherent contribution''). In mesoscopic optics, the latter process is responsible for the well known coherent backscattering effect \cite{Aegerter09}. In the present context, both diagrams (b) and (c) contribute exactly the same amount to fluctuations. Their evaluation is however more involved than that of diagram (a) because of the two additional Green tensors connecting the scattering processes at $\br_1$ and $\br_2$. The main lines of the derivation are presented in Appendix \ref{diagramb_calc} for clarity. The final result for the double scattering contribution to fluctuations, $\av{\delta U^2_{(2)}}(z_A)$, reads
\begin{eqnarray}
\label{deltaU2_final}
&&\av{\delta U^2_{(2)}}(z_A)=
\dfrac{\hbar^2}{c^{12}}
\int_0^\infty
\dfrac{\rmd \xi_1}{2\pi}
\dfrac{\rmd \xi_2}{2\pi}
\dfrac{\rmd^2 \bq_a}{(2\pi)^2}
\dfrac{\rmd^2 \bq_b}{(2\pi)^2}
\dfrac{\rmd^2 \bq_d}{(2\pi)^2}
\xi_1^6\xi_2^6\nonumber\\
&&\times
\alpha(i\xi_1)\alpha(i\xi_2)
\dfrac{e^{-(\kappa_a+\kappa_b+\kappa_c+\kappa_d)z_A}}{8 \kappa_a\kappa_b\kappa_c\kappa_d}
\dfrac{n^2\alpha_s^4}{4\pi(\kappa_a+\kappa_b+\kappa_c+\kappa_d)}\nonumber\\
&&\times
\int \rmd\hat{\br}
\sum_{p_a, p_b, p_c, p_d}
\left(\bs{\epsilon}_a\cdot\bs{\epsilon}_b\right)^2
\left(\bs{\epsilon}_c\cdot\bs{\epsilon}_d\right)^2
\nonumber\\
&&
\times\text{Re}\bigg\{
[\bs{\epsilon}_a\cdot\bs{\epsilon}_b-(\bs{\epsilon}_a\cdot\hat{\br})(\bs{\epsilon}_b\cdot\hat{\br})]
[\bs{\epsilon}_c\cdot\bs{\epsilon}_d-(\bs{\epsilon}_c\cdot\hat{\br})(\bs{\epsilon}_d\cdot\hat{\br})]
\nonumber\\
&&\times\left[-i\hat{\br}\cdot(\bq_b-\bq_d)|\hat{\br}\times\hat{\bs{z}}|+
|\hat{\br}\cdot\hat{\bs{z}}|\dfrac{\kappa_a+\kappa_b+\kappa_c+\kappa_d}{2}\right.\nonumber\\
&&
+\left.
\hat{\br}\cdot\hat{\bs{z}}\dfrac{\kappa_a+\kappa_c-\kappa_b-\kappa_d}{2}+\dfrac{\xi_1+\xi_2}{c}
\right]^{-1}
\bigg\},
\end{eqnarray}
where the central integral refers to an average over the direction of $\hat{\br}$. 
As in sec. \ref{results} we introduce the ratio
\begin{equation}
\label{gamma2_def}
\gamma_{(2)}\equiv
\dfrac{\sqrt{\av{\delta U^2_{(2)}}(z_A)}}{|\av{U}(z_A)|}
=\dfrac{\sqrt{\av{\delta U^2_{(2)}}(z_A)}}{\eta |U_*|},
\end{equation}
which measures the contribution of double scattering to relative fluctuations. From eq. (\ref{deltaU2_final}), we find the following asymptotic limits:
\begin{equation}
\label{gamma2}
\gamma_{(2)}=
     \begin{cases}
      \dfrac{a_2 n\alpha_s}{n z_A^3} & z_A\gg\lambda_A \\
      \dfrac{b_2 n\alpha_s}{n z_A^2\lambda_A} & z_A\ll\lambda_A,
     \end{cases}
\end{equation}
with $a_2\simeq 0.15$ and $b_2\simeq 0.43$. In fig. \ref{fluc_double} we show $\gamma_{(2)}$ as a function of $z_A/\lambda_A$, in units of the dimensionless parameter $n\alpha_s/(n\lambda_A^3)$. Unlike $\gamma$, $\gamma_{(2)}$ is not a function of the single parameter $nz_A^3$, as is clear from fig. \ref{fluc_double} and eq. (\ref{gamma2}).
Furthermore, as compared to eq. (\ref{gamma}), the double scattering contribution (\ref{gamma2}) comes with an additional factor $n\alpha_s\sim n a^3\ll 1$. In other words, for a dilute repartition of heterogeneities, double scattering is negligible compared to single scattering. The situation would of course be different for denser disorder such that $n\alpha_s\sim1$. In this limit, double, and presumably all higher multiple scattering processes, become of the same order of magnitude as single scattering and must be accounted for in the estimation of fluctuations.

As a final comment, we mention that in our approach we have neglected a number of scattering processes where an individual wave path is scattered more than once by the same scatterer (``recurrent scattering'') \cite{Wiersma95}. Fig. \ref{diagrams}(d) shows such a process as an example. In practice, recurrent scattering is negligible in dilute disordered media, but may be significant for denser disorder. A detailed treatment of recurrent scattering would require to modify both the effective dielectric constant $\tilde{\epsilon}$ and the correlator $U(\br_1-\br_1')$, a task far beyond the scope of this paper.

\section{Conclusion}

We have calculated the statistical fluctuations of the Ca-simir interaction potential between a two-level atom and a disordered material, in the limit of no interface reflection. For a dilute distribution of identical, independent Rayleigh scatterers, our results indicate that these fluctuations are dominated by non-specular reflection on a single scatterer. The relative fluctuations of the Casimir potential at a distance $z_A$ from the medium are then inversely proportional to the square root of the number of scatterers in a volume $z_A^3$. This demonstrates that fluctuations cannot be neglected when the atom-surface distance becomes smaller than the average distance between the scatterers. These results are consistent with previous work concerned with the classical Casimir force induced by thermal fluctuations at high temperatures \cite{Dean10}, and with recent works on the Casimir effect in metals \cite{Alloca15}. They additionally specify the conditions of validity of the study presented in \cite{Dufour13b}. From a practical point of view, our study could explain the surprisingly low values of atomic quantum reflection observed in recent experiments, which could be attributed to atoms reflected non specularly on the fluctuations of the Casimir potential \cite{Pasquini06}. In a similar context, such decrease of quantum reflection due to heterogeneities might constitute a limitation of the ability of nano-porous materials to efficiently trap or guide antimatter \cite{Dufour13b}.

Although we have focused on the case of independent, Rayleigh scatterers, our approach can be applied to more general situations where the scatterers are not point like or where they are spatially correlated (via a modification of the average dielectric constant $\tilde\epsilon$ and of the correlation function $U$). With minor changes, our theory can also account for finite optical thickness of the medium or for internal reflections at the interface with the vacuum. Finally, it could in principle also be used to describe materials having high concentrations of heterogeneities by calculating the full multiple scattering (``ladder'') series, albeit this is likely to be a difficult task \cite{Muller02}.


\section*{Acknowledgements}

The authors thank Dominique Delande and Cord M\"uller for insightful discussions.

\appendix

\section{Scalar products of polarization vectors}
\label{scalar}

In this appendix, we provide explicit expressions for the scalar products of polarization vectors that appear in eq. (\ref{deltaU1_final}).
Polarization vectors are defined as
\begin{eqnarray}
&&\be_\text{TE}(\bq_i)=\hat{\bs{z}}\times\hat{\bq}_i, \nonumber\\
&&\be_\text{TM}(\bq_i)=\be_\text{TE}(\bq_i)\times\hat{\bk_i}.
\end{eqnarray}
where $\bk_i=\bq_i+ k_i^z\hat{\bs{z}}$ for $i=a, c$ (incoming modes), and $\bk_i=\bq_i- k_i^z\hat{\bs{z}}$ for $i=b, d$ (outgoing modes), with $k_i^z=\sqrt{\xi_1^2/c^2+\bq_i^2}$ for $i=a, b$, $k_i^z=\sqrt{\xi_2^2/c^2+\bq_i^2}$ for $i=c, d$, and $\bq_c=\bq_a-\bq_b+\bq_d$. eq. (\ref{deltaU1_final}) involves integrals over (i) the angle $\phi$ between $\bq_b$ and $\bq_d$, (ii) the angle $\phi'$ between $\bq_a$ and $\bq_d$, and (iii) $q_a\equiv|\bq_a|$, $q_b\equiv|\bq_b|$ and $q_d\equiv|\bq_d|$. With these definitions, we show in Tables \ref{table1} and \ref{table2} the scalar products $\be_a\cdot\be_b\equiv \be_{p_a}(\bq_a)\cdot\be_{p_b}(\bq_b)$ and $\be_c\cdot\be_d\equiv \be_{p_c}(\bq_c)\cdot\be_{p_d}(\bq_d)$, respectively.

\begin{table}[h]
\begin{tabular}{>{\centering}p{1.1cm}|>{\centering}p{2.6cm}|>{\centering}p{3.7cm}}
&
$p_b=$TE
&
$p_b=$TM
\tabularnewline
\hline
$p_a=$TE
&
$\cos(\phi'-\phi)$
&
$\dfrac{c\kappa_b\sin(\phi'-\phi)}{\xi_1}$
\tabularnewline
\hline
$p_a=$TM
&
$\dfrac{c\kappa_a\sin(\phi'-\phi)}{\xi_1}$&
$-c^2\dfrac{q_a q_b+\kappa_a\kappa_b\cos(\phi'-\phi)}{\xi_1^2}$
\tabularnewline
\end{tabular}
\caption{\label{table1} Values of $\be_{p_a}(\bq_a)\cdot \be_{p_b}(\bq_b)$ for the various combinations of polarizations.}
\end{table}
\begin{table}[h]
\begin{tabular}{>{\centering}p{1.cm}|>{\centering}p{3.2cm}|>{\centering}p{3.5cm}}
&
$p_d=$TE
&
$p_d=$TM
\tabularnewline
\hline
$p_c=$TE
&
$\dfrac{q_d-q_b\cos\phi+q_a\cos\phi'}{q_c}$
&
$\dfrac{c\kappa_d}{\xi_2}\dfrac{q_a\sin\phi'-q_b\sin\phi}{q_c}$
\tabularnewline
\hline
$p_c=$TM
&
$\dfrac{c\kappa_c}{\xi_2}\dfrac{q_a\sin\phi'-q_b\sin\phi}{q_c}$
&
$-\bigg[\dfrac{q_d-q_b\cos\phi+q_a\cos\phi'}{q_c}$\\
$\times \kappa_c\kappa_d+q_c q_d)\bigg]\dfrac{c^2}{\xi^2_2}$
\tabularnewline
\end{tabular}
\caption{\label{table2} Values of $\be_{p_c}(\bq_c)\cdot \be_{p_d}(\bq_d)$ for the various combinations of polarizations. Here $q_c\equiv|\bq_a-\bq_b+\bq_d|$ $=\sqrt{q_a^2+q_b^2+q_d^2-2 q_a q_b\cos(\phi'-\phi)+2q_a q_d\cos\phi'-2 q_b q_d\cos\phi}$.}
\end{table}

\section{Double scattering correlation function}
\label{diagramb_calc}

In this appendix, we give the main steps that lead to eq. (\ref{deltaU2_final}). We here focus on the calculation of the diagram in fig. \ref{diagrams}(b) (diagram(c) is calculated analogously, and gives the same final result). The mathematical formulation of the diagram in fig. \ref{diagrams}(b) is

\begin{eqnarray}
\label{rrdouble}
&&\av{\delta r_{ab}(\omega_1)\delta r_{cd}^*(\omega_2)}^{(2)}
=4 k_a^z k_c^{z*}
\int
\rmd^2\bs{\rho}_a
\rmd^2\bs{\rho}_b
\rmd^2\bs{\rho}_c
\rmd^2\bs{\rho}_d
\nonumber\\
&&
(nk_1^2 k_2^2\alpha_s^2)^2
\int_{z_1>0}\!\!\!\!\!\!\!\rmd^3 \br_1
\int_{z_2>0}\!\!\!\!\!\!\!\rmd^3 \br_2
e^{i (\bq_a\cdot\bs{\rho}_a-\bq_b\cdot\bs{\rho}_b-\bq_c\cdot\bs{\rho}_c+\bq_d\cdot\bs{\rho}_d)}\nonumber\\
&&
\left[\be_a
\cdot\av{\bs{G}}_{\omega_1}(\{\bs{\rho}_a,0\},\br_1)
\cdot\av{\bs{G}}_{\omega_1}(\br_1, \br_2)
\cdot\av{\bs{G}}_{\omega_1}(\br_2, \{\bs{\rho}_b,0\})
\cdot \be_b
\right]\nonumber\\
&&
\left[\be_c^*\cdot\av{\bs{G}}_{\omega_2}^*(\{\bs{\rho}_c,0\},\br_1)
\cdot\av{\bs{G}}_{\omega_2}^*(\br_1, \br_2)
\cdot\av{\bs{G}}_{\omega_2}^*(\br_2, \{\bs{\rho}_d,0\})\cdot \be_d^*
\right],\nonumber\\
\end{eqnarray}
where we have already performed two spatial integrations, making use of the Dirac delta form of the correlator $U$. We now change the variables from $(\br_1,\br_2)\equiv$ $(\{\bs{\rho}_1, z_1\}, \{\bs{\rho}_2, z_2\})$ to $(\{\bs{R}=(\bs{\rho}_1+\bs{\rho}_2)/2, z_1\}, \{\bs{\rho}=\bs{\rho}_1-\bs{\rho}_2, z_2\})$, use eq. (\ref{rr_intermediate}) for the four Green tensors connected to the interface, and perform integrations over $\bs{\rho}_a$, $\bs{\rho}_b$, $\bs{\rho}_c$, $\bs{\rho}_d$ and $\bs{R}$. This yields
\begin{eqnarray}
&&\av{\delta r_{ab}(\omega_1)\delta r_{cd}^*(\omega_2)}^{(2)}
=(nk_1^2 k_2^2\alpha_s^2)^2
\dfrac{(2\pi)^2\delta(\Delta\bq_a-\Delta\bq_b)}{4 k_b^z k_d^{z*}}\nonumber\\
&&\times\intdones{\bs{\rho}}\int_0^\infty{dz_1}\int_0^\infty dz_2
\ e^{i(k_a^z-k_c^{z*})z_1+i(k_b^z-k_d^{z*})z_2+i\bs{\rho}\cdot\bs{\Delta q}_a}\nonumber\\
&&\times \left[\be_a\cdot\av{\bs{G}}_{\omega_1}(\bs{\rho},z_1, z_2)\cdot\be_b\right]
\otimes \left[\be_c\cdot\av{\bs{G}}_{\omega_2}(\bs{\rho}, z_1, z_2)\cdot\be_d\right]^*.
\end{eqnarray}
Making use of eq. (\ref{Gbar_inf}) and neglecting near-field contributions, we approximate the first term within square brackets as
\begin{equation}
-\dfrac{e^{i k_1r}}{4\pi r}
\left[\be_a\cdot\be_b-(\be_a\cdot\hat{\br}_1)(\be_b\cdot\hat{\br}_2)
\right],
\end{equation}
where $\br\equiv\{\bs{\rho}, z_1-z_2\}$, and similarly for the second term. 
As in the calculation of the single scattering contribution, we replace all wave numbers $\tilde{k}_i^z$ by $k_i^z$, and $\tilde{k}_i$ by $k_i$, which is a good approximation in the dilute limit (\ref{diluteness}).
We then introduce the new change of variables $(z_1,z_2)\rightarrow(z=z_1+z_2,z_{12}=z_1-z_2)$ and perform the integral over $z$. We obtain
\begin{eqnarray}
\label{appendix_final}
&&\av{\delta r_{ab}(\omega_1)\delta r_{cd}^*(\omega_2)}^{(2)}
=(nk_1^2 k_2^2\alpha_s^2)^2
\dfrac{(2\pi)^2\delta^{(2)}(\Delta\bq_a-\Delta\bq_b)}{4 k_b^z k_d^{z*}}\nonumber\\
&&\times\int \rmd^3{\br}
\ e^{i(k_a^z-k_c^{z*}+k_b^z-k_d^{z*})|z_{12}|/2
+i(k_a^z-k_c^{z*}-k_b^z+k_d^{z*})z_{12}/2}\nonumber\\
&&\times \left[\be_a\cdot\be_b-(\be_a\cdot \hat{\br})(\be_b\cdot \hat{\br})\right]\left[\be_c^*\cdot\be_d^*-(\be_c^*\cdot \hat{\br})(\be_d^*\cdot \hat{\br})\right]\nonumber\\
&&\times e^{i\br\cdot\bs{\Delta q}_a} \dfrac{e^{i (k_1-k_2^*)r}}{(4\pi r)^2}
\dfrac{i}{k_a^z-k_c^{z*}+k_b^z-k_d^{z*}},
\end{eqnarray}
with the definition $\int \rmd^3\br\equiv\int \rmd^2\bs{\rho}\int_{-\infty}^{\infty} dz_{12}$. eq. (\ref{deltaU2_final}) of the main text is finally obtained by performing the integration over $r=|\br|$, inserting eq. (\ref{appendix_final}) into eq. (\ref{deltaUdef}), and applying the double Wick rotation as explained in the main text.


\begin{thebibliography}{99}
\bibitem{footnote}
When squaring eq. (\ref{Udef}), one also generates contributions involving the correlator $\av{\delta r_{ab}(\omega_1)\delta r_{cd}(\omega_2)}$. These contributions can be reduced to the form (\ref{deltaUdef}) by making use of the causality of reflection coefficients.
\end{thebibliography}
\end{document}